\documentclass[aps,prl,twocolumn]{revtex4}
\usepackage{graphicx}
\usepackage{latexsym}
\usepackage{amsmath}
\usepackage{amsfonts}
\usepackage{amssymb}
\usepackage{color}
\usepackage{units}
\usepackage{natbib}
\newcommand{\be}{\begin{equation}}
\newcommand{\ee}{\end{equation}}
\newcommand{\bea}{\begin{eqnarray}}
\newcommand{\eea}{\end{eqnarray}}
\usepackage[parfill]{parskip}
\usepackage{orcidlink}
\begin{document}
\title{Quorum sensing with density-enhanced motility and size regulation} 
\author{Itay Azizi \orcidlink{0000-0003-2939-4421}}
\affiliation{Independent Researcher, Vitoria Gasteiz, Spain}
\email{itay.azizi@gmail.com} 
\begin{abstract} Motivated by biological systems and my previous studies of quorum sensing with density-enhanced motility \cite{Azizi2026A,Azizi2026B}, I study a model of density-enhanced size and motility, in which both particle size and motility increase when the local density (over a distance larger than particle diameter) exceeds the global density. The emergent structures along different size ratios are spots, holes, bands and labyrinths controlled by several distinct observations. First, the characteristic pattern length scale is controlled by the sensing radius and remains approximately independent of the size ratio. Second, the total area occupied by the particles decreases with increasing size ratio. Third, the active fraction barely depends on the size ratio. These results demonstrate how both particle size and motility can couple microscopic regulation to large-scale pattern formation. I conclude by discussing possible extensions of the model.

\end{abstract}
\maketitle
\section{Introduction}
To illustrate density-dependent behavior, consider a system of \(N\) shoppers inside a supermarket. When the local density of shoppers is low, they wander randomly and independently. However, when a shopper senses, either by sight or sound, more than \(X\) other shoppers within a distance \(Y\), their behavior changes. Here, \(N\) sets the global density, \(X\) defines the quorum size, and \(Y\) defines the sensing range. In addition, there are interactions between the shoppers and the specific form of sensing. After some time, the interplay between these three parameters, individual behavior, and collective regulation can lead to steady patterns with phase separation and characteristic morphologies.

The supermarket scenario illustrates quorum sensing, a universal mechanism underlying social interactions within groups and therefore observed across diverse disciplines. Recent theoretical studies have considered toy models in which motility depends discontinuously on the local density, with motility either correlated or anti-correlated with density \cite{Bauerle2018,Azizi2026A,Azizi2026B}. These studies demonstrate that density-dependent activity can generate a variety of emergent structures, including circular clusters, holes, stripes, and labyrinths.

Particle size is also a parameter that can depend on density. For example, body size can depend on population density across species. In the sea urchin Diadema antillarum, individuals become smaller at high densities and larger at low densities \cite{levitan1989life}. In California sea lions, body size increases during population recovery, a change partly attributed to intensified sexual selection favoring larger males \cite{Valenzuela2023}. In herbivores, increasing density can reduce forage quality and consequently individual body mass \cite{HOBBS2024}. Human studies further demonstrate that the relationship between population density and stature is context dependent, with both positive and negative associations reported across populations \cite{Foster1983,Komlos2007,Nikoi2013,Zhang2019}.

Spatial organization as observed in quorum sensing of density-dependent motility is also widespread in animal collectives. Juvenile locusts, for example, form large collective structures known as hopper bands. These groups can adopt different morphologies, including broad bands, narrow ribbons, compact spots, and, in some cases, intertwined streams or networks. Such structures emerge from local interactions between individuals rather than from centralized coordination \cite{Dkhili2017}. Fish schools provide another example in which local interactions can generate large-scale collective organization. 

Thus, increasing local density may be associated with individuals becoming smaller, becoming larger, or maintaining their size, depending on the biological system and the underlying mechanisms. This diversity motivates the study of size regulation as a generic mechanism for collective organization.

Here, I investigate the collective behavior of coupling local density, particle size and activity at long-range quorum sensing. Specifically, I ask how density-enhanced motility and size regulation modifies the emergent morphology of a quorum-sensing system. What macroscopic structures emerge? How does density-dependent size regulation affect the resulting area fraction?
\begin{figure}[ht] 
\includegraphics[width=0.7\linewidth]{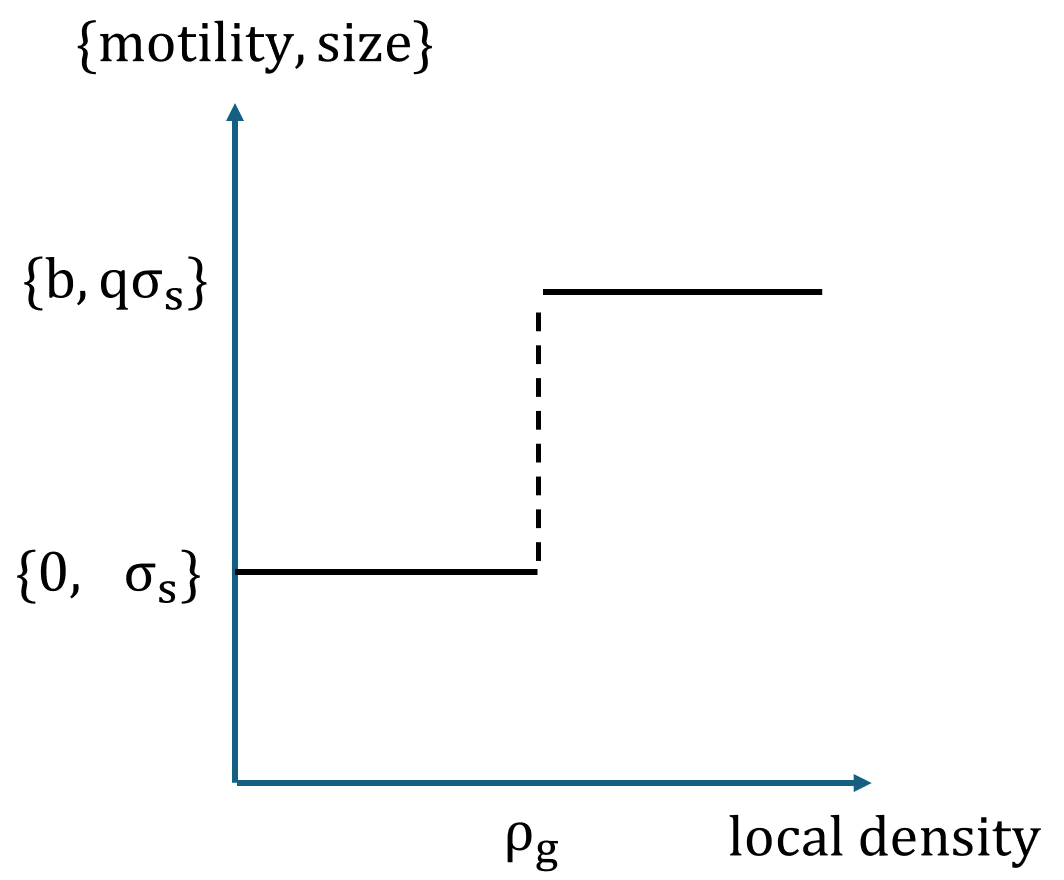}
	\caption{\label{fig:1} Density-enhanced motility and size regulation. When the local density exceeds a critical value equal to the global density, $\rho_g$, the particle increases both its motility, from $Pe=0$ to $Pe=b$, and its size, from $\sigma_s$ to $q\sigma_s$.}
\end{figure}
\section{Methods}
In order to elucidate the behavior of systems exhibiting density-enhanced motility and size, Langevin dynamics simulations were carried out using a code written in Fortran. The two-dimensional system consists of $N=4000$ particles in a square box with periodic boundary conditions and dimensions $L=L_x=L_y=100\sigma$ with $\sigma$ as the unit of length which sets a global density of $\rho_g=0.4\sigma^{-2}$. 
\\Particles $i$ and $j$ interact via a Weeks-Chandler-Andersen short range repulsive potential 
\begin{equation}
V(r_{ij})=4\epsilon[(\sigma_{ij}/r_{ij})^{12}-(\sigma_{ij}r_{ij})^{6}]+\epsilon
\end{equation}
with $\epsilon=1$, cutoff distance of $r_c=2^{1/6}\sigma_{ij}$, $r_{ij}=\lvert\mathbf{r}_j-\mathbf{r}_i\rvert$ and $\sigma_{ij}=(\sigma_i+\sigma_j)/2$.  The choice of a purely repulsive potential ensures that any observed clustering behavior can be attributed to quorum sensing rather than to direct attractive interactions. I set $k_B=1$ and $T=0.02$; such a low temperature yields approximately hard-core interactions with $2^{1/6}\sigma_{ij}$ approximately equal to the particle diameter.

Particle $i$ alternates between "passive+small" and "active+big" identities as a function of its local density $\rho_i$. The code computes $m_i$, the number of neighbors of particle $i$ within a disk of radius $R_{sense}$ centered on particle $i$; thus the local density is $\rho_i = m_i/\pi R_{sense}^2$. The global and critical densities are taken to be equal, so that the particle responds to deviations from the ambient density $\rho_g$ (see Fig.\ref{fig:1}).

For a given size ratio $q>1$, a small particles is assigned with size parameter of $\sigma_s$, a big  particle with $\sigma_b=q\sigma_s$ and the small-big interaction with $\sigma_{sb}=(\sigma_s+\sigma_b)/2$. 

For an equimolar (50:50) mixture, the mean size is normalized to $<\sigma_i>=1$. Thus, $\frac{\sigma_s+q\sigma_s}{2}=1$ and 
\begin{equation}
\begin{gathered}
\sigma_s=\frac{2}{1+q}  
\\ \sigma_b=\frac{2q}{1+q}
\end{gathered}
\end{equation}
I consider increasing size ratios, $q=1.05$, $1.22$, $1.50$ and $1.80$. These values span a range from weak to pronounced size polydispersity. The smallest ratio, $q=1.05$, represents a relatively weak size contrast, comparable in magnitude to the sexual difference in human height. The intermediate ratio, $q=1.22$, represents a substantially stronger size contrast, relevant to animal species exhibiting pronounced sexual size dimorphism in body length or height. The largest ratio, $q=1.50$, corresponds, for approximately spherical particles of equal density, to a volume ratio of $q^3=3.375$. Thus, the largest size ratio introduces a substantial contrast in particle volume and mass, while the smaller ratios probe progressively weaker forms of size polydispersity. Overall, the selected range provides a physically meaningful progression from near-monodisperse to strongly polydisperse binary systems, while remaining relevant to size differences observed across biological species. These biological comparisons are intended only to provide an intuitive scale for the size ratios; the particle model does not reproduce the detailed geometry, physiology, or scaling relationships of biological organisms.

Particle $i$ is assigned a position $r_i$ and an orientation $\theta_i$, which evolve according to the following Langevin equations:
\begin{equation}
\begin{gathered}
\frac{d\mathbf{r}_i(t)}{dt}=v_p(\rho_i) \mathbf{e_i}(t) - \beta D \nabla U_i(t)+\sqrt{2D}\boldsymbol{\eta}(t) 
\\\frac{d\theta_i(t)}{dt}=\sqrt{2D_r(\rho_i)}\xi(t)
\end{gathered}
\end{equation}

where $v_p$ is the magnitude of the self-propulsion velocity, oriented along $\mathbf{e}_i(t) = (\cos\theta_i(t), \sin\theta_i(t))$. The Péclet number ($\text{Pe}$) is defined as the ratio of the persistence length to rotational diffusion length: 
\begin{eqnarray}
Pe=\dfrac{l_p}{l_D} =\dfrac{v_p}{\sqrt{DD_r}}.
\end{eqnarray}
\\The translational (rotational) diffusion coefficient is denoted by $D$ ($D_r$), and $\boldsymbol{\eta}(t)$ and $\xi(t)$ are Gaussian white noise terms satisfying $\langle \eta_i(t) \eta_j(t^{\prime}) \rangle = \delta_{ij}\delta(t-t^{\prime})$, with $i,j \in (x,y)$, and $\langle \xi(t)\xi(t^{\prime}) \rangle = \delta(t-t^{\prime})$, respectively. $\gamma$ and $\gamma_r$ denot{e the translational and rotational friction coefficients, satisfying $D = k_BT/\gamma$ and $D_r = k_BT/\gamma_r$, respectively, with $\gamma = 2$ and $\gamma_r = 0.01$.
$U$ denotes the potential energy arising from conservative forces, e.g., pair interactions such that $U_i = \sum_{j \neq i} V(r_{ij})$, where the gradient is taken with respect to the position of particle $i$. 
According to local density, a "small+passive particle" is assigned with $\sigma_s$ and $v_p = \text{Pe} = D_r = 0$, while a "big+active" particle is assigned with $\sigma_b=q\sigma_s$ and $v_p$ such that $\text{Pe} = b$.
The constants for all systems studied are $\rho, T, \gamma, \gamma_r$, denoting the density, temperature, translational friction coefficient, and rotational friction coefficient, respectively. The variables are $R_{sense},q$, denoting the sensing radius and size ratio, respectively. Thus, each simulation studies a single point in this $(R_{sense},q)$ phase space.
\\Each simulation begins from a square lattice and is run for up to $10^5$ steps with a timestep of $\Delta t = 0.005$, which is sufficient for relaxation, as monitored by plateaus in the potential energy and small particle fraction, and by small variation in cluster size and morphology. 

\section{Results}

\subsection{Visual inspection}

Hereby, I will describe my understanding of the system using the visual data.
The snapshots of the systems in steady state are shown in Figs.\ref{fig:2}-\ref{fig:3} in a state diagram as a function of size ratio and activity. At zero activity there are two cases. At low sensing radius $R_{sense}=5\sigma$, across all size ratios sampled, the small and big particles are partially mixed and there are small domains of big particle aggregates. At high sensing radius $R_{sense}=15\sigma$, there are large big particle aggregates and as size ratio increases, the big particle domains shift gradually from a large cluster to a network structure of small particles with spots of big particles. 

At low activity corresponds to $b=10$ for low sensing radius $R_{sense}=5\sigma$, there is a network of small particles with a small fraction of big particles and noticeble voids. This tendencty is preserved for all size ratios checked. For the high sensing radius $R_{sense}=15\sigma$, there are elongated bands of small particles with peristaltic deformations and dilute gas of big particles. As size ratio increases, the bands get connected. 

At high activity, corresponding to $b=40$, the morphology remains qualitatively the same as for $b=10$. However, the clusters contain a larger fraction of small particles, making them more compact, while the gas phase contains fewer big particles. In addition, the cluster interfaces become flatter.

\begin{figure}[ht] 
\includegraphics[width=0.95\linewidth]{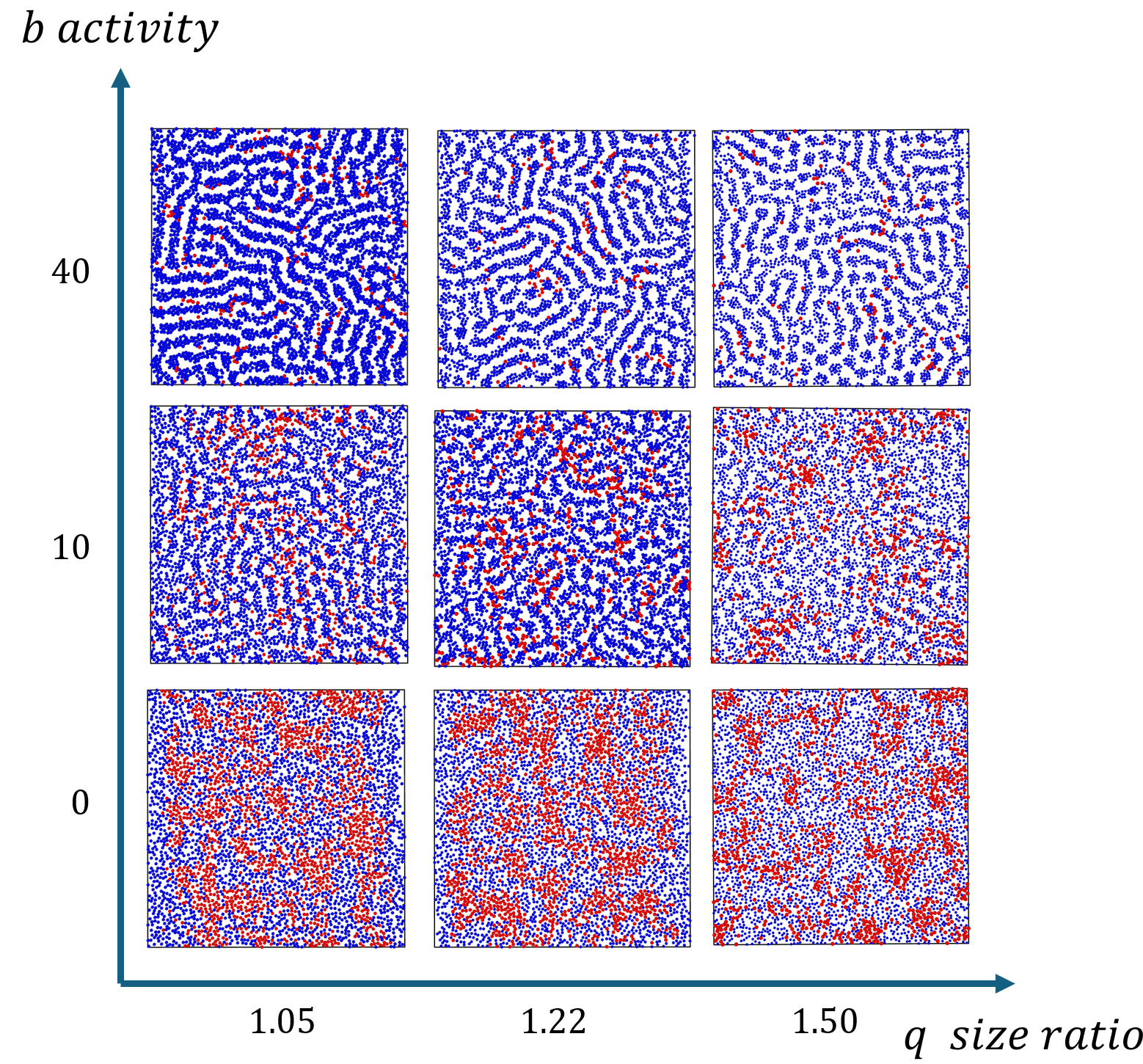}
	\caption{\label{fig:2} Snapshots of steady states at a small sensing radius, $R_{sense}=5\sigma$, as a function of the size ratio $q$ and activity $b$. Big particles are shown in red, while small particles are shown in blue. Increasing activity leads to increasingly pronounced pattern formation, characterized by labyrinthine structures formed by the small particles. As activity increases, the active fraction decreases.}
\end{figure}

Typically, the domain width $w$, defined as the thickness of the bands, is approximately two-thirds of the sensing radius:
\begin{equation}
w \approx \frac{2}{3}R_{\mathrm{sense}}.
\end{equation}

For a fixed sensing radius, changing the size ratio therefore does not directly change the characteristic domain size. Instead, smaller particles allow a larger number of particles to occupy a domain of the same characteristic size. This larger number of particles can modify the local density fluctuations and the dynamic interplay between satisfying the quorum-sensing condition and maintaining dynamic heterogeneity between the slow and fast components.

\begin{figure}[ht] 
\includegraphics[width=0.95\linewidth]{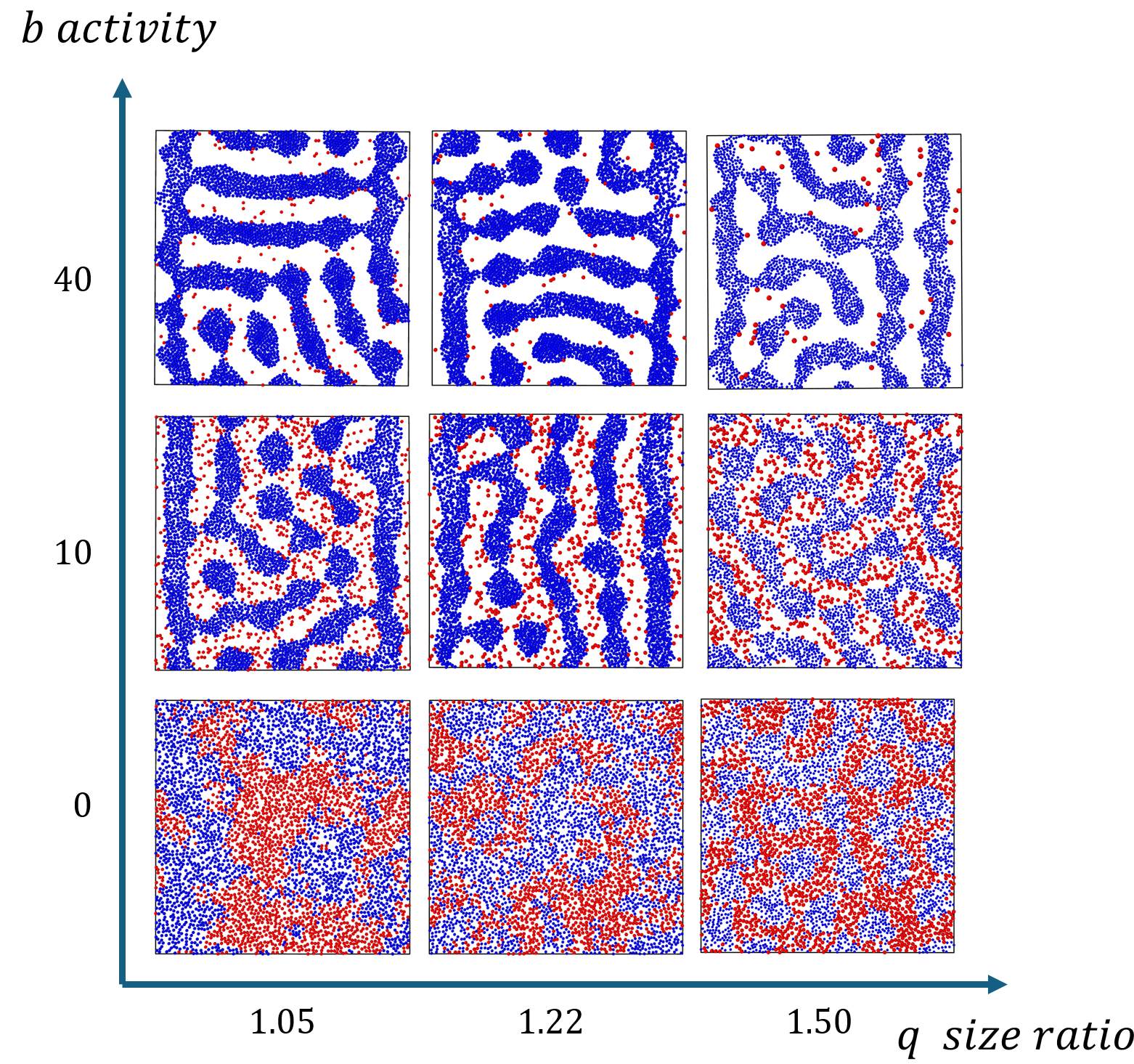}
	\caption{\label{fig:3} 
    Snapshots of steady states at a large sensing radius, $R_{sense}=15\sigma$, showing the dependence of the resulting patterns on the size ratio $q$ and activity $b$. Big particles are shown in red, while small particles are shown in blue. Increasing activity enhances pattern formation by the small particles, leading to labyrinthine structures composed of bands with peristaltic deformations. As activity increases, the active fraction decreases.}
\end{figure}

\subsection{Steady-state observables}

 Steady state averages of area fraction $\phi_{area}$ and active fraction $\phi_{active}$ are presented in Fig. \ref{fig:4} and Fig.\ref{fig:5}, respectively. The area fraction is defined as the ratio of the total area occupied by the mixture of small and big particles to the area occupied by a monodisperse system of the same number of particles with a single size parameter $\sigma_i=\sigma$: 
\begin{equation}
\phi_{\mathrm{area}} = \frac{N_s  \sigma_s^2 + N_b \sigma_b^2}{N\sigma^2}.
\end{equation}
Here, $N_s$ and $N_b$ denote the numbers of small and big particles, respectively, while $\sigma_s$ and $\sigma_b$ denote their corresponding size parameters. The area fraction characterizes the spatial occupancy of the group that may have a functional or biological significance. 

For $b=10$, the area fraction decreases as a function of size ratio (see Fig.\ref{fig:4}(i)). This means that the clusters formed by small particles maintain similar size across different size ratios. Since the small particles in the larger area fractions are smaller, then the total area occupied by the particles is smaller. This tendency is similar for the two sensing ranges presented. 
For $b=40$, the area fraction also decreases as a function of size ratio from the same reasons. However it decreases faster than $b=10$, which agrees with the visual observations. 

\begin{figure}[ht] 
\includegraphics[width=0.67\linewidth]{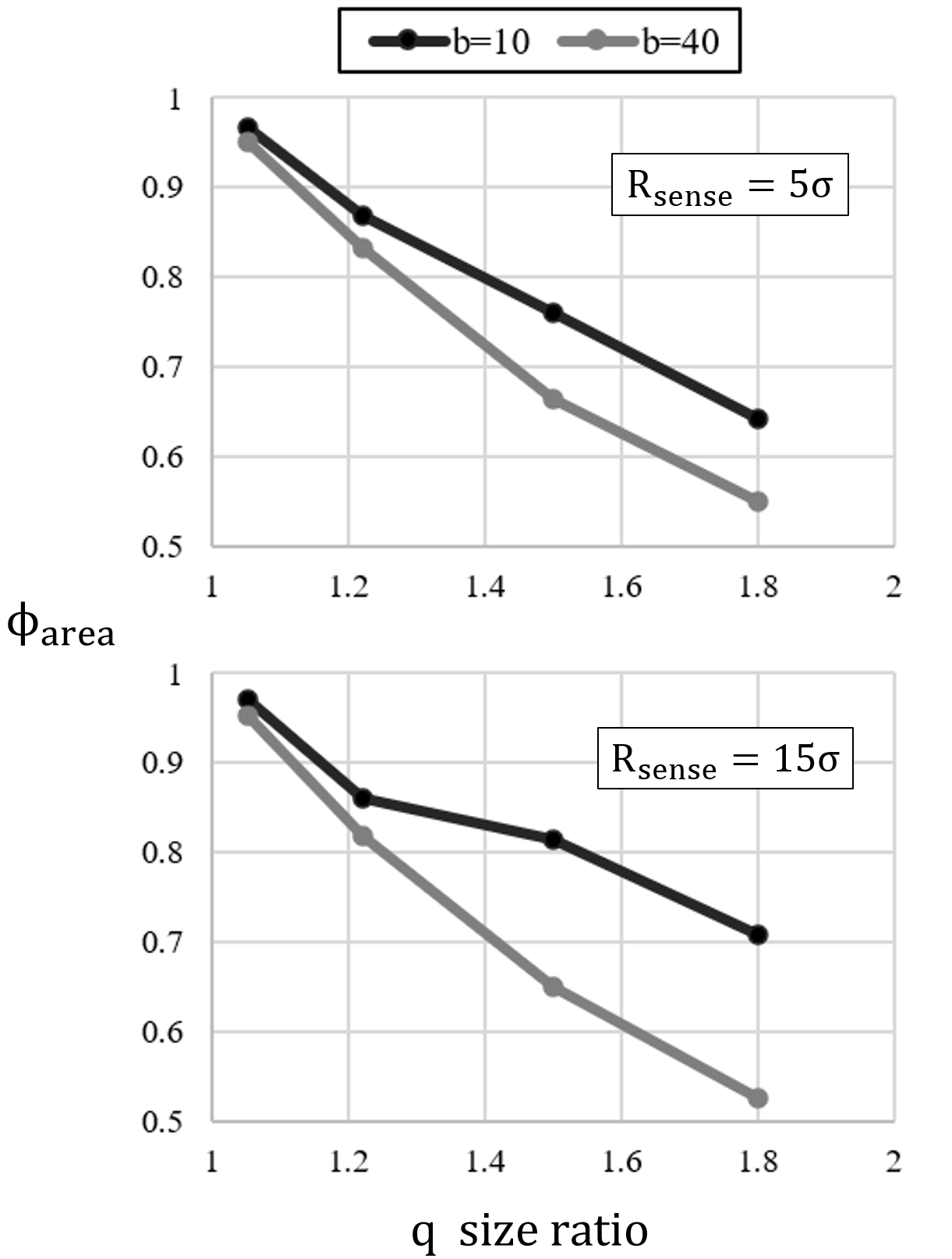}
	\caption{\label{fig:4} Steady-state area fraction, $\phi_{\mathrm{area}}$, as a function of the size ratio $q$ for different values of activity and sensing radius. For $q$ close to unity, the area fraction is slightly below unity. At high activity, the area fraction decreases significantly, reaching values as low as $\phi_{area}\approx 0.55$.}
\end{figure}

The active fraction is defined as the ratio of the number of active particles to the total number of particles:
\begin{equation}
\phi_{\mathrm{active}} = \frac{N_{\mathrm{active}}}{N}.
\end{equation}
This parameter characterizes the degree to which the system contains fast components. 

The overall trend of active fraction is decreasing with size ratio, with some regions showing monotonic behavior (see Fig.\ref{fig:5}). At low activity, the active fraction is of order 0.1, while at high activity, it decreases to order of 0.01. This suggests that activity strongly promoting pattern formation of small particles  on the expanse of sampling big particles. 

\begin{figure}[ht] 
\includegraphics[width=0.67\linewidth]{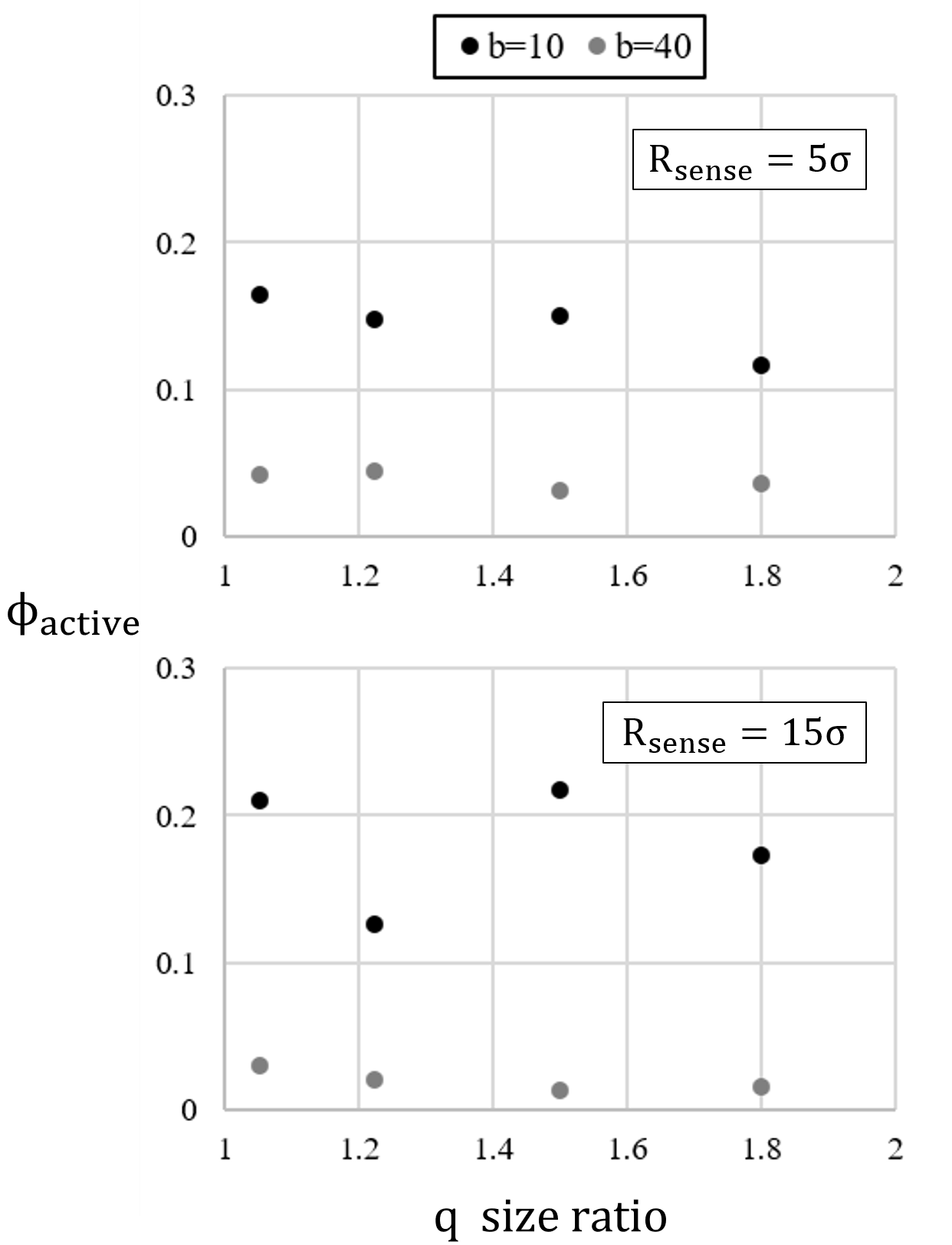}
	\caption{\label{fig:5} Steady-state active fraction as a function of the size ratio $q$ for different values of activity and sensing radius. The active fraction changes mostly as a function of activity, but barely changes as a function of size ratio}
\end{figure}

\begin{figure}[ht] 
\includegraphics[width=0.95\linewidth]{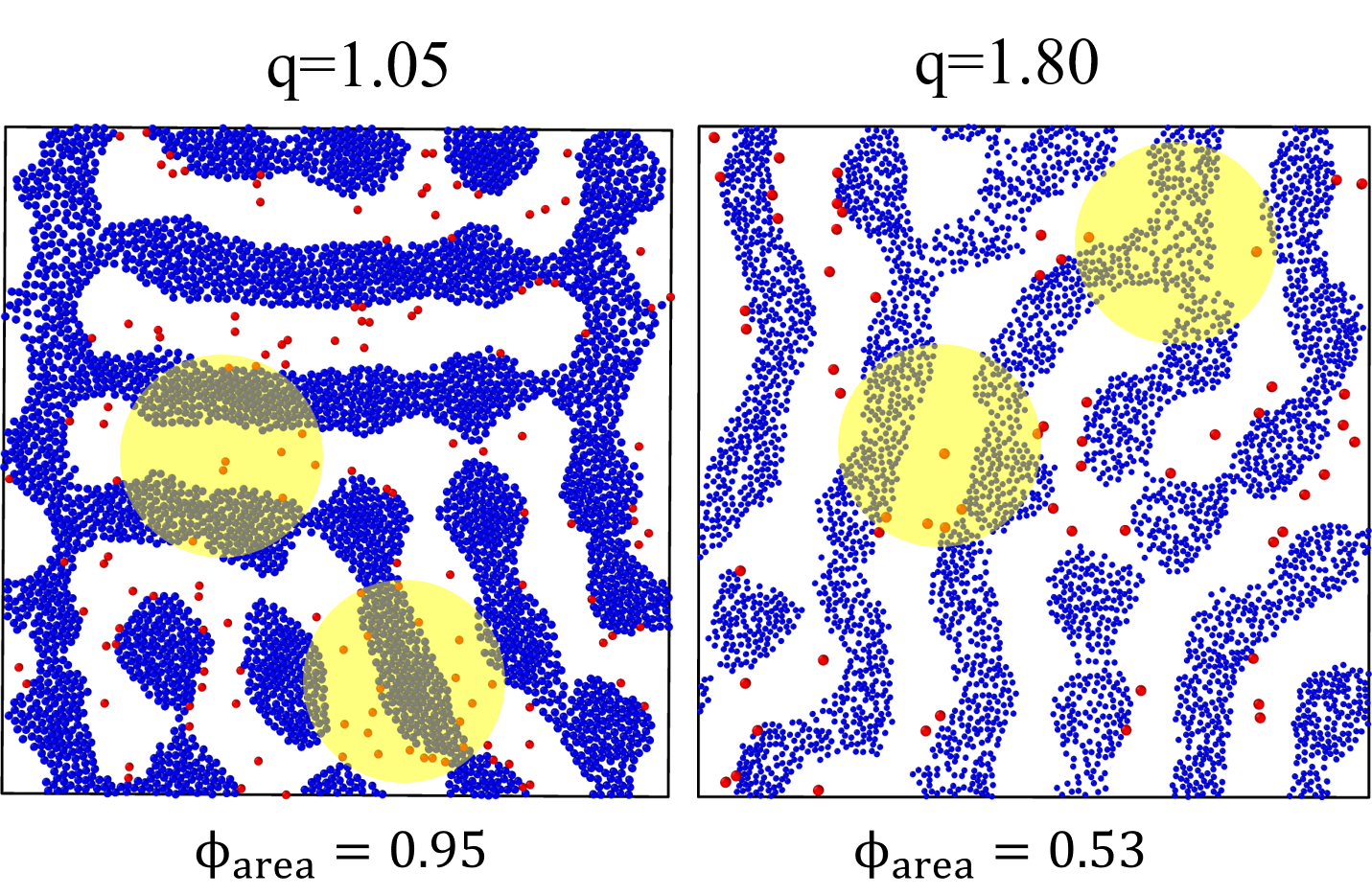}
	\caption{\label{fig:6} Comparison between two strikingly different emerging strucurues at $R_{sense}=15\sigma$ and $b=40$. Big particles are shown in red, small particles are shown in blue and examples for sensing windows are shown in yellow.}
\end{figure}

When comparing two strikingly different emerging structures at $R_{sense}=15\sigma$ and $b=40$, the bands have the same characteristic length scale (see Fig.~\ref{fig:6}). However, their internal structures are different. At $q=1.8$, the clusters are dense, whereas at $q=1.05$, they are more fluid-like.

\section{Summary}

Size and energy polydispersity in equilibrium can lead to the formation of glasses and mosaic crystals \cite{Azizi2020}. The present work, however, is not in equilibrium and the interplay between particle size and sensing length scale results in formation of morphologies, such as spots, holes, bands and labyrinths. There are three main observations in this work. 
\begin{enumerate}
\item \textbf{Pattern-size invariance:} Despite changes in the particle size ratio, the shape and characteristic length scale of the emergent patterns remain approximately unchanged. This observation supports my finding in \cite{Azizi2026B} that the sensing radius is the dominant parameter controlling the pattern length scale.
\item \textbf{Area reduction:} As the size ratio increases, the formation of clusters of small particles leads to a global reorganization that reduces the total area occupied by the collective structure. This mechanism may have functional implications for biological or ecological systems in which collective structures benefit from occupying less space.
\item \textbf{Activity:} The active fraction depends mostly on activity and sensing radius but not on size ratio. This indicates that the fraction of highly mobile particles is barely affected by the selection of sizes.
\end{enumerate}

\section{Discussion}

Several directions could be pursued to extend the present model. First, the framework could be generalized to three dimensions, allowing the emergence and stability of three-dimensional structures to be investigated.

Second, the model could be extended to explicitly incorporate sexual dimorphism, in which females retain a fixed size while males can increase or decrease their size in response to the local density (see Fig.~\ref{fig:7}). This extension could first be studied in passive systems and subsequently generalized to active particles, with the effects of sexual attraction and repulsion either included or neglected. Such a model would provide a simple framework for investigating how density-dependent size regulation interacts with sex-specific interactions and motility.

\begin{figure}[ht] 
\includegraphics[width=0.90\linewidth]{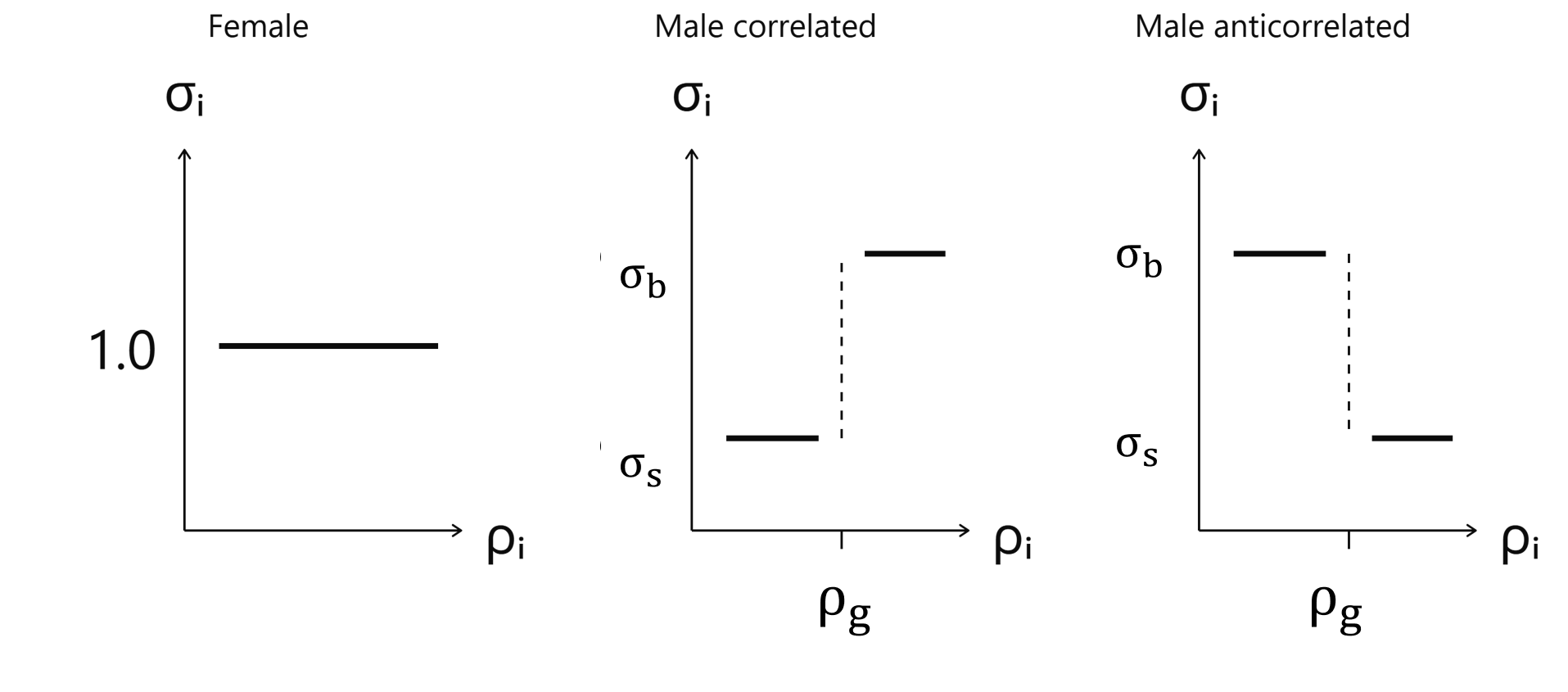}
	\caption{\label{fig:7} Toy model of density-dependent sexual dimorphism in which the sizes of the two sexes respond differently to local density. While females retain their size independent of local density, male can increase or decrease their size as local density crosses the global density of the system.}
\end{figure}
Third, the size and motility responses could be controlled by independent critical densities. In the present formulation, both responses are triggered by the same density threshold; relaxing this constraint would allow $\rho_{\mathrm{size}}$ and $\rho_{\mathrm{motility}}$ to vary independently, such that $\rho_{\mathrm{size}} \neq \rho_{\mathrm{motility}}$. This additional degree of freedom could lead to new regimes in which size regulation precedes motility enhancement, or vice versa, potentially giving rise to further morphological transitions.

The study of quorum sensing as a generic model of social interaction is still in its early stages. I expect this interdisciplinary field to expand significantly, creating connections between previously separate disciplines and bringing together their distinct perspectives, methods, and knowledge to identify common principles of social interaction and collective behavior.

\section{Acknowledgments}
The author thanks Oren Pedatzur for valuable scientific discussions and feedback. This work received no external financial support and was entirely self-funded.
\bibliography{references}
\end{document}